\documentclass[reprint,aps,amsmath,amssymb,floatfix,longbibliography,prb]{revtex4-2}

\usepackage{xcolor}
\usepackage{graphicx}

\usepackage{bm}
\usepackage{dcolumn}

\usepackage{amsmath}

\usepackage[
colorlinks=true,
urlcolor=blue,
citecolor=blue,
linkcolor=blue,
hyperfootnotes=false]{hyperref}

\newcommand{\bk}{\mathbf k}
\newcommand{\bq}{\mathbf q}

\newcommand{\bQ}{\mathbf Q}
\newcommand{\br}{\mathbf r}

\begin{document}
\title{Antisymmetric Dynamical Spin Correlations:\\
Spin-Space-Group Constraints and Frequency-Moment Sum Rules}
\author{Tsutomu Momoi}
\affiliation{RIKEN Center for Emergent Matter Science, Wako, Saitama, 351-0198, Japan}

\date{\today}

\begin{abstract}

Polarized inelastic neutron scattering probes the handedness of magnetic excitations 
through the spin-component-antisymmetric part of the dynamical spin-structure-factor tensor.
We derive its symmetry constraints under unitary and antiunitary
residual operations of magnetic space groups and spin-space groups.
These constraints determine the allowed tensor components, their parity under momentum reversal, 
and symmetry-enforced nodes.
We then introduce the corresponding antisymmetric commutator spectral function 
and derive exact zero-temperature sum rules for its frequency moments.
The zeroth moment is fixed by the uniform magnetization.
For Heisenberg or XXZ exchange Hamiltonians,
the first moment of the $xy$ component is fixed by a momentum-weighted static vector chirality.
For compensated collinear antiferromagnets, the antisymmetric dynamical structure factor
is forbidden when inversion or a translation relates the opposite-spin sublattices.
In rotation-related altermagnets, however, an $xy$ component that is even under momentum reversal
is allowed at generic wave vectors.
Finite antisymmetric spectral weight can therefore coexist
with vanishing zeroth and first moments of the commutator spectral function.
In a planar helix, by contrast, a residual antiunitary spin-space-group
symmetry makes the allowed $xy$ component odd under momentum reversal.
Its first moment is generally nonzero and is fixed by a momentum-weighted static vector chirality
through the first-moment sum rule.
Linear spin-wave calculations for representative altermagnetic and helical models explicitly realize these symmetry and sum-rule constraints.


\end{abstract}
\maketitle

\section{Introduction}

Polarized neutron scattering probes both the diagonal components of the dynamical spin-correlation tensor
and the imaginary parts of its off-diagonal components~\cite{Blume1963,Maleev2002}.
In the Blume--Maleev formalism, correlations that are antisymmetric in the spin components
contribute to the polarization-dependent part of the magnetic neutron-scattering cross section,
commonly called the chiral term.
In the elastic channel, this term probes
the handedness of static noncollinear magnetic structures and is closely related to
their static vector chirality~\cite{PlumerKawamuraCaille1991,Seki2008,Janoschek2010,Simonet2012}.
Here we instead consider the inelastic channel, in which the chiral term probes the
spin-component-antisymmetric part of the dynamical spin-correlation tensor.
This contribution is sensitive to the circular
polarization, and hence the handedness, of magnetic excitations and can be finite 
even when the static vector chirality vanishes.

The spin-component-antisymmetric dynamical correlations 
have been studied in several classes of magnetic systems. 
In ferromagnets and ferrimagnets, the inelastic chiral signal has been used to identify the
polarization and handedness of magnon excitations~\cite{Nambu2020,Jenni2022}.
In noncollinear magnets, including helical magnets,
dynamical chirality is tied to the handedness of the ordered
spin texture and has been studied both theoretically and
experimentally in triangular antiferromagnets, chiral langasites, and
MnSi~\cite{Syromyatnikov2005,Loire2011,Weber2019}.
More recently, collinear
altermagnets~\cite{Naka2019,Smejkal2022PRX} have provided a distinct
setting.
In the compensated collinear limit considered here, both the uniform
magnetization and the static vector chirality vanish.
Theoretical studies have investigated magnon splitting in altermagnets~\cite{Naka2019,Smejkal2023ChiralMagnons}
and its signatures in neutron scattering~\cite{MaierOkamoto2023,McClarty2025}.
Chiral magnon splitting has been observed by neutron
scattering in MnTe~\cite{Liu2024MnTe,Liu2026MnTe}.

Taken together, these examples show that a finite spin-component-antisymmetric dynamical
signal can originate from uniform magnetization in a ferromagnet, vector chirality
in a helical magnet, or magnon-band splitting in an altermagnet.
This variety motivates a systematic framework that determines when
the antisymmetric dynamical response is allowed and distinguishes
among its different microscopic origins.

In this paper, we develop such a framework by combining
symmetry constraints on antisymmetric dynamical correlations with
moment sum rules for the corresponding commutator spectral function.
The central quantity is the
spin-component-antisymmetric dynamical structure factor
\(C^{\alpha\beta}(\mathbf q,\omega)\), which we call the
antisymmetric DSF.
We first derive the constraints imposed on this quantity by unitary and antiunitary
operations of magnetic space groups and spin-space groups.
For a time-reversal-invariant state, the antiunitary constraint
recovers the behavior discussed by Maleev for a paramagnet:
the chiral response is odd under momentum reversal~\cite{Maleev2002}.
More generally, the residual symmetries of an ordered state determine
the allowed tensor components, their parity under momentum reversal, and
symmetry-enforced nodes.

To complement these symmetry constraints, we introduce the
corresponding antisymmetric commutator spectral function and derive
exact sum rules for its zeroth and first frequency moments.
The zeroth moment is fixed by the uniform magnetization.
For models with Heisenberg or XXZ exchange, the first moment of the
\(xy\) component is fixed by a momentum-weighted static vector
chirality.
Together, the momentum-parity constraints and moment sum rules 
distinguish among antisymmetric dynamical responses arising from
uniform magnetization, static vector chirality, or magnon-band
splitting.

We apply this framework to collinear antiferromagnets and planar
helical magnets.
For the collinear antiferromagnets considered here, the
spin-space-group classification of Chen {\it et al.}~\cite{Chen2025SSGMagnons}
provides a compact way to organize the selection rules.
The antisymmetric DSF is forbidden when inversion or a translation relates the opposite-spin
sublattices. 
By contrast, in rotation-related altermagnets, an \(xy\) component that is even under momentum reversal 
can remain finite at generic wave vectors.
In a minimal altermagnetic model, this response arises from split
magnon branches with opposite circular polarizations, although both the
zeroth and first moments of the commutator spectral function vanish.
A planar helix provides a contrasting case: a residual antiunitary
spin-space-group symmetry makes the $xy$ component odd under momentum reversal.
Its first moment is allowed by symmetry and is generally nonzero;
the first-moment sum rule fixes it in terms of the static vector chirality.

The remainder of this paper is organized as follows.
Section~II defines the spin-component-antisymmetric dynamical structure factor.
Sections~III and IV derive its symmetry constraints and frequency-moment sum rules, respectively.
Sections~V and VI apply this framework to collinear
antiferromagnets and planar helical magnets, respectively.
Section~VII summarizes the results, and Appendices~A and B provide details of the
linear spin-wave calculations.

\section{Spin-component-antisymmetric dynamical structure factor}

We define the zero-temperature dynamical spin-structure-factor tensor as
\begin{equation}
S^{\alpha\beta}(\bq,\omega)
=
\sum_n
\langle 0|
\delta S_{-\bq}^{\alpha}
|n\rangle
\langle n|
\delta S_{\bq}^{\beta}
|0\rangle
\delta(\omega-E_n+E_0),
\end{equation}
where
\[
\delta S^\alpha_{\bq}
=
S^\alpha_{\bq}
-
\langle 0| S^\alpha_{\bq} |0\rangle,\qquad
S^\alpha_{\bq}
=
\frac{1}{\sqrt{N}}
\sum_i e^{-i\bq\cdot\mathbf r_i}S_i^\alpha.
\]
Here \(S_i^\alpha\) is the \(\alpha\) component of the spin operator
at site \(i\), \(\mathbf r_i\) is the position of that site,
\(\alpha,\beta\in\{x,y,z\}\), and \(N\) is the number of sites.
The states \(|n\rangle\) form a complete set of eigenstates of the
Hamiltonian with energies \(E_n\), and \(|0\rangle\) denotes the
ground state.

We focus on the part of this tensor that is antisymmetric under interchange of the spin components:
\begin{equation}
C^{\alpha\beta}(\bq,\omega)
\equiv
\frac{1}{2i}
\left[
S^{\alpha\beta}(\bq,\omega)
-
S^{\beta\alpha}(\bq,\omega)
\right].
\end{equation}
We call \(C^{\alpha\beta}(\bq,\omega)\) the spin-component-antisymmetric
dynamical structure factor, or simply the antisymmetric DSF.
Hermiticity gives
$\delta S^\alpha_{\bq}
=
(\delta S^\alpha_{-\bq})^\dagger$. Consequently,
$S^{\beta\alpha}(\bq,\omega)
=
\left[
S^{\alpha\beta}(\bq,\omega)
\right]^*$ and 
\begin{equation}
C^{\alpha\beta}(\bq,\omega)
=
\operatorname{Im} S^{\alpha\beta}(\bq,\omega).
\end{equation}
In the Blume--Maleev formalism, this antisymmetric DSF enters the chiral term of 
the polarized-neutron-scattering cross section~\cite{Blume1963,Maleev2002}.

\section{Symmetry constraints}
\label{sec:symmetry_constraints}

We derive symmetry constraints on the spin-component-antisymmetric DSF
using standard selection-rule arguments for tensor-valued correlation
functions~\cite{Birss1964,InternationalTablesD2014}.

\subsection{General constraints}

We first formulate these constraints in a form that applies to both magnetic
space groups and spin-space groups.
The dynamical spin-structure-factor tensor can be written as
\begin{equation}
S^{\alpha\beta}(\bq,\omega)
=
\left\langle
\delta S^\alpha_{-\bq}
\delta(\omega-H+E_0)
\delta S^\beta_{\bq}
\right\rangle ,
\label{eq:dsf_definition_symmetry}
\end{equation}
where
\(\langle O\rangle\equiv \langle 0|O|0\rangle\).
We work within a particular ordered ground state corresponding to a single magnetic domain
and retain only the residual symmetry operations that leave it invariant.

We characterize a residual symmetry operation by its real-space part
\(g=\{R|\mathbf t\}\), written in Seitz notation, and by an orthogonal
matrix \(\Lambda\) acting on spin components:
\begin{align*}
g: \mathbf{r} \mapsto R \mathbf{r} + \mathbf{t},\qquad
\Lambda: S^\alpha_i \mapsto \Lambda_{\alpha\gamma} S_i^\gamma.
\end{align*}
Here and below, repeated spin indices are summed over.
We leave \(\Lambda\) unspecified so that the same notation applies to
magnetic space groups and spin-space groups.
In a magnetic space group, \(\Lambda\) is fixed by the real-space operation,
whereas in a spin-space group the spin rotation is an
independent part of the symmetry operation.
We specify the corresponding forms of \(\Lambda\) in Secs.~\ref{sec:MSG} and \ref{sec:SSG}.
For an antiunitary operation \(\Theta=\theta U\), \(\Lambda\)
denotes only the spin-space action of the unitary part \(U\);
the spin reversal under time reversal \(\theta\) is treated explicitly below.

For a unitary residual symmetry \(U\) with real-space part
\(g=\{R|\mathbf t\}\) and spin-space matrix \(\Lambda\),
the Fourier spin operator transforms as
\begin{equation}
U S^\alpha_{\bq}U^{-1}
=
e^{i(R\bq)\cdot\mathbf t}
\Lambda_{\alpha\gamma}
S^\gamma_{R\bq}.
\label{eq:unified_unitary_spin_operator}
\end{equation}
The fluctuation operators \(\delta S^\alpha_{\bq}\) transform in the same way.
In Eq.~\eqref{eq:dsf_definition_symmetry}, the phase factors associated with the two
spin operators cancel.
Consequently, the antisymmetric DSF obeys
\begin{equation}
C^{\alpha\beta}(\bq,\omega)
=
\Lambda_{\alpha\gamma}
\Lambda_{\beta\delta}
C^{\gamma\delta}(R\bq,\omega).
\label{eq:unified_unitary_C_constraint}
\end{equation}

For an antiunitary residual symmetry, we write
\(\Theta=\theta U\), where \(U\) has real-space part
\(g=\{R|\mathbf t\}\) and spin-space matrix \(\Lambda\).
The Fourier spin operator transforms as
\begin{equation}
\Theta S^\alpha_{\bq}\Theta^{-1}
=
-
e^{-i(R\bq)\cdot\mathbf t}
\Lambda_{\alpha\gamma}
S^\gamma_{-R\bq}.
\label{eq:unified_antiunitary_spin_operator}
\end{equation}
The minus sign comes from spin reversal under time reversal,
and antiunitarity complex conjugates the phase factor.
In the two-point function, the spin-reversal signs and phase factors
cancel, whereas antiunitarity complex conjugates the correlation function:
\begin{equation}
S^{\alpha\beta}(\bq,\omega)
=
\Lambda_{\alpha\gamma}
\Lambda_{\beta\delta}
\left[
S^{\gamma\delta}(-R\bq,\omega)
\right]^* .
\label{eq:unified_antiunitary_S_constraint}
\end{equation}
Using the identity
$\left[
S^{\gamma\delta}(\bq,\omega)
\right]^*
=S^{\delta\gamma}(\bq,\omega)$,
we obtain
\begin{equation}
C^{\alpha\beta}(\bq,\omega)
=
-
\Lambda_{\alpha\gamma}
\Lambda_{\beta\delta}
C^{\gamma\delta}(-R\bq,\omega).
\label{eq:unified_antiunitary_C_constraint}
\end{equation}
The overall minus sign thus follows directly from the complex
conjugation required by antiunitarity.

\subsection{Magnetic space groups}\label{sec:MSG}

We first specialize the general constraints to magnetic space groups.
Because spin is an axial vector, a real-space point operation $R$ acts on it as
\begin{equation}
\mathbf S
\mapsto
D(R)\mathbf S,
\qquad
D(R)=\det(R)R .
\label{eq:msg_axial_vector_matrix}
\end{equation}
Consequently, for an operation of a magnetic space group with real-space part \(g=\{R|\mathbf t\}\),
the matrix \(\Lambda\) in the general constraints is
\begin{equation}
\Lambda=D(R).
\label{eq:msg_Lambda_definition}
\end{equation}
For an antiunitary magnetic operation, the additional spin reversal under time reversal is
treated separately, as in Eq.~\eqref{eq:unified_antiunitary_C_constraint}.

\subsection{Spin-space groups}\label{sec:SSG}

We next specialize the general constraints to spin-space groups (SSGs).
SSGs describe symmetries for which spin rotations need not
be tied to real-space point operations~\cite{BrinkmanElliott1966,LitvinOpechowski1974}.
We denote unitary and antiunitary SSG operations by
\begin{equation}
(R_s\Vert g)\quad \text{and}\quad
(\theta R_s\Vert g),
\qquad
g=\{R|\mathbf t\},
\label{eq:ssg_operation_definition}
\end{equation}
where \(R_s\) acts in spin space and \(g\) acts in real space.
In the general constraints, the spin-space matrix is therefore
\begin{equation}
\Lambda=R_s.
\label{eq:ssg_Lambda_definition}
\end{equation}

\subsection{Corollaries for spin-space-group symmetries}
\label{sec:corol}

We collect several consequences of the general constraints that will
be used in the applications below.  We focus on spin-space-group
symmetries, where the spin and real-space parts of a symmetry
operation need not be tied to one another.

Here and below, a condition such as
\(R\bq=\bq\) means equality of the physical momentum vectors,
not merely equality modulo a reciprocal-lattice vector \(\mathbf G\). 
Under our Fourier convention, which uses the actual site
positions, \(S^\alpha_{\bq+\mathbf G}\) generally differs from
\(S^\alpha_{\bq}\) in a crystal with multiple sites per primitive
cell because of basis-dependent phase factors.

\paragraph{Unitary spin-\(\pi\)-rotation symmetries.}

Consider a unitary spin-space-group operation
\((C_{2x}^s\Vert g)\),
where \(C_{2x}^s\) denotes a spin \(\pi\) rotation
about the \(x\) axis and \(g=\{R|\mathbf t\}\).
The spin rotation \(C^s_{2x}\) is represented by \(\Lambda=\mathrm{diag}(1,-1,-1)\),
so Eq.~\eqref{eq:unified_unitary_C_constraint} gives
\[
C^{x\mu}(\bq,\omega)
=
-
C^{x\mu}(R\bq,\omega),
\qquad
\mu=y,z .
\]
Thus, \(C^{x\mu}\) vanishes whenever
\(R\bq=\bq\).

For a pure translation
\((C_{2x}^s\Vert T_{\mathbf t})\), one has \(R=E\), and therefore
\[
C^{x\mu}(\bq,\omega)=0,
\qquad
\mu=y,z .
\]
Similarly, a symmetry \((C_{2y}^s\Vert T_{\mathbf t'})\) gives
\[
C^{y\nu}(\bq,\omega)=0,
\qquad
\nu=x,z .
\]
If two orthogonal spin-\(\pi\)-rotation symmetries with pure-translation 
real-space parts are present, all off-diagonal components of the
antisymmetric DSF are forbidden:
\[
C^{\alpha\beta}(\bq,\omega)=0,
\qquad
\alpha\ne\beta .
\]

\paragraph{Antiunitary symmetry selection rule.}

Consider an antiunitary spin-space-group operation
\((\theta R_s\Vert g)\), with \(g=\{R|\mathbf t\}\).
Equation~\eqref{eq:unified_antiunitary_C_constraint} gives
\[
C^{\alpha\beta}(\bq,\omega)
=
-
(R_s)_{\alpha\gamma}
(R_s)_{\beta\delta}
C^{\gamma\delta}(-R\bq,\omega).
\]
In general, this relation may mix different pairs of spin components.
A useful special case occurs when the tensor action of $R_s$ maps
a given component to itself up to a sign \(\eta_{\alpha\beta}=\pm1\).
The constraint then reduces to
\[
C^{\alpha\beta}(\bq,\omega)
=
-\eta_{\alpha\beta}
C^{\alpha\beta}(-R\bq,\omega),
\]
where no summation over \(\alpha\) and \(\beta\) is implied.

A component with \(\eta_{\alpha\beta}=+1\) therefore vanishes at any momentum 
satisfying $-R\bq=\bq$.
If this condition holds throughout the Brillouin zone, the component
is forbidden at all momenta.
If it holds only at a high-symmetry point or along a high-symmetry line or plane, 
the component has a symmetry-enforced node there.

The simplest example of a prohibition throughout the Brillouin zone is
\((\theta\Vert\{P|\mathbf0\})\), where \(P\) denotes spatial inversion.
In this case, \(R=-E\) and
the unitary spin-space part is trivial, \(R_s=E\), so that
\(\eta_{\alpha\beta}=+1\).
Because \(-R\bq=\bq\) for every \(\bq\), this symmetry forbids all off-diagonal components 
of the antisymmetric DSF:
\[
C^{\alpha\beta}(\bq,\omega)=0,
\qquad
\alpha\ne\beta,
\]
for all $\bq$ and $\omega$.

\section{Sum rules for the antisymmetric commutator spectral function}
\label{sec:sum_rules}

The symmetry constraints derived in the previous section determine
which components of \(C^{\alpha\beta}(\bq,\omega)\) are allowed and
how they transform in momentum space.  
To complement these constraints, we 
first derive an equal-time relation for the
energy-integrated antisymmetric DSF and then establish exact frequency-moment sum rules
for the corresponding full-frequency commutator spectral function.

\subsection{Energy-integrated antisymmetric DSF}
At zero temperature, integrating the ordinary DSF over positive frequencies gives
\begin{equation}
 \int_{0}^{\infty} d\omega\,
 S^{\alpha\beta}(\bq,\omega)
 =
 \left\langle
 \delta S_{-\bq}^{\alpha}\delta S_{\bq}^{\beta}
 \right\rangle .
\label{eq:ordinary_DSF_zeroth_moment}
\end{equation}
It follows that
\begin{equation}
 \int_{0}^{\infty} d\omega\,
 C^{\alpha\beta}(\bq,\omega)
 =
 \frac{1}{2i}
 \left\langle
 \delta S_{-\bq}^{\alpha}\delta S_{\bq}^{\beta}
 -
 \delta S_{-\bq}^{\beta}\delta S_{\bq}^{\alpha}
 \right\rangle .
\label{eq:ordinary_C_zeroth_moment}
\end{equation}
The right-hand side is the part of the equal-time connected
correlation function that is antisymmetric under interchange of the spin components.
For a cyclic permutation \((\alpha,\beta,\gamma)\) of \((x,y,z)\),
this quantity can be decomposed into the Fourier sine transform of the 
$\gamma$ component of the connected intersite vector chirality,
\(\langle\delta\mathbf S_i\times\delta\mathbf S_j\rangle^\gamma\), and
an onsite contribution proportional to the uniform magnetization.
Thus, Eq.~\eqref{eq:ordinary_C_zeroth_moment} gives the equal-time
content of the energy-integrated inelastic chiral signal.
This positive-frequency integral generally depends on momentum;
the commutator zeroth moment introduced below selects its
momentum-even combination.

\subsection{Commutator spectral function}

Maleev formulated the polarization-dependent inelastic chiral
response in terms of the spin-component-antisymmetric part of a
retarded commutator Green function~\cite{Maleyev1995,Maleev2002}.
Here we instead work directly with the associated full-frequency commutator spectral function
and derive exact sum rules from its moments.

At zero temperature, 
\(S^{\alpha\beta}(\bq,\omega)\) has support only at nonnegative frequencies.
We define the associated full-frequency commutator
spectral function by
\begin{align}
\mathcal A^{\alpha\beta}(\bq,\omega)
& \equiv
S^{\alpha\beta}(\bq,\omega)
-
S^{\beta\alpha}(-\bq,-\omega).
\end{align}
Its spin-component-antisymmetric part is
\begin{equation}
\mathcal A_{\rm as}^{\alpha\beta}(\bq,\omega)
\equiv
\frac{1}{2i}
\left[
\mathcal A^{\alpha\beta}(\bq,\omega)
-
\mathcal A^{\beta\alpha}(\bq,\omega)
\right].
\end{equation}
Using the definition of \(C^{\alpha\beta}(\bq,\omega)\), we obtain 
\begin{equation}\label{eq:Aas_C_relation}
\mathcal A_{\rm as}^{\alpha\beta}(\bq,\omega)
=
C^{\alpha\beta}(\bq,\omega)
+
C^{\alpha\beta}(-\bq,-\omega).
\end{equation}
Thus, \(\mathcal A^{\alpha\beta}_{\rm as}\) is the full-frequency extension of the
antisymmetric DSF \(C^{\alpha\beta}\).

We define the zeroth and first full-frequency moments of \(\mathcal A^{\alpha\beta}_{\rm as}\) by
\begin{equation}
M_n^{\alpha\beta}(\bq)
\equiv
\int_{-\infty}^{\infty}
d\omega\,\omega^n
\mathcal A_{\rm as}^{\alpha\beta}(\bq,\omega),
\qquad n=0,1 .
\label{eq:moment_definition_01}
\end{equation}
Equation~\eqref{eq:Aas_C_relation} then gives the moment relation
\begin{equation}
M^{\alpha\beta}_n(\bq)
=
\int_0^\infty d\omega\,\omega^n
\left[
C^{\alpha\beta}(\bq,\omega)
+
(-1)^n C^{\alpha\beta}(-\bq,\omega)
\right].
\label{eq:C_positive_frequency_moments}
\end{equation}
Thus, the zeroth and first moments select the momentum-even and momentum-odd parts of
\(C^{\alpha\beta}\), respectively.

\subsection{Zeroth moment}

Integrating the commutator spectral function over frequency gives
\[
\int_{-\infty}^{\infty}
d\omega\,
\mathcal A^{\alpha\beta}(\bq,\omega)
=
\left\langle
[
S^\alpha_{-\bq},
S^\beta_{\bq}
]
\right\rangle .
\]
The fluctuation subtraction does not affect this result because
\(c\)-number expectation values drop out of the commutator.
Taking the spin-component-antisymmetric part gives
\begin{equation}
M^{\alpha\beta}_{0}(\bq) = \int_{-\infty}^{\infty}
d\omega\,
\mathcal A_{\rm as}^{\alpha\beta}(\bq,\omega)
=
\frac{\epsilon_{\alpha\beta\gamma}}{N} \sum_i
\left\langle S_i^\gamma \right\rangle.
\label{eq:zeroth_commutator_sum_rule}
\end{equation}
Here \(\epsilon_{\alpha\beta\gamma}\) is the Levi-Civita symbol, with
\(\epsilon_{xyz}=+1\).
Equivalently,
\begin{equation}
\int_0^\infty d\omega\,
\left[
C^{\alpha\beta}(\bq,\omega)
+
C^{\alpha\beta}(-\bq,\omega)
\right]
=
\frac{\epsilon_{\alpha\beta\gamma}}{N}
\sum_i
\langle S_i^\gamma\rangle .
\label{eq:C_zeroth_sum_rule}
\end{equation}
Equation~\eqref{eq:C_zeroth_sum_rule} shows that the
zeroth moment, which probes the momentum-even part of
\(C^{\alpha\beta}(\bq,\omega)\), is fixed by the uniform
magnetization.
Specifically, \(M^{\alpha\beta}_0\) is independent of \(\bq\).
For a cyclic permutation \((\alpha,\beta,\gamma)\) of \((x,y,z)\), 
it equals the \(\gamma\) component of the uniform magnetization
density.
In particular, \(M^{xy}_0\) measures the magnetization density
along the \(z\) axis.

\subsection{First moment}

The first moment of the commutator spectral function obeys the
standard double-commutator sum rule~\cite{HohenbergBrinkman1974}
\begin{equation}
\int_{-\infty}^{\infty} d\omega\, \omega\, \mathcal A^{\alpha\beta}(\bq,\omega)
=
\left\langle
\left[
S^\alpha_{-\bq},
[H,S^\beta_{\bq}]
\right]
\right\rangle .
\end{equation}
Taking the spin-component-antisymmetric part gives
\begin{align}
M_1^{\alpha\beta}(\bq) &= \int_{-\infty}^{\infty} d\omega\, \omega\, \mathcal A_{\rm as}^{\alpha\beta}(\bq,\omega) \nonumber\\
&=
\frac{1}{2i}
\left(\left\langle
\left[
S^\alpha_{-\bq},
[H,S^\beta_{\bq}]
\right] \right\rangle
-
\left\langle \left[
S^\beta_{-\bq},
[H,S^\alpha_{\bq}]
\right]
\right\rangle \right) .
\label{eq:1sum_A}
\end{align}

As an example, consider the XXZ Hamiltonian
\begin{equation}
H_{\rm XXZ}
=
\sum_{\langle ij\rangle}
\left[
J^\perp_{ij}
(S_i^xS_j^x+S_i^yS_j^y)
+
J^z_{ij}S_i^zS_j^z
\right].
\end{equation}
For the \(xy\) component, evaluating the two double commutators gives
\begin{align}
M_{1}^{xy}(\bq)&=\int_{-\infty}^{\infty} d\omega\, \omega\, \mathcal A_{\rm as}^{xy}(\bq,\omega)\nonumber\\
&=
\frac{1}{N}
\sum_{\langle ij\rangle}
J^z_{ij}
\sin\!\left[
\bq\cdot(\mathbf r_i-\mathbf r_j)
\right]
\left\langle
(\mathbf S_i\times\mathbf S_j)^z
\right\rangle .
\label{eq:1st_sum_rule}
\end{align}
Equivalently,
\begin{align}
&
\int_0^\infty d\omega\,\omega
\left[
C^{xy}(\bq,\omega)-C^{xy}(-\bq,\omega)
\right]
\nonumber\\
&\hspace{1cm}
=
\frac{1}{N}
\sum_{\langle ij\rangle}
J^z_{ij}
\sin[\bq\cdot(\mathbf r_i-\mathbf r_j)]
\left\langle
(\mathbf S_i\times\mathbf S_j)^z
\right\rangle .
\label{eq:positive_frequency_1st_sum_rule}
\end{align}
For the XXZ Hamiltonian, Eq.~\eqref{eq:positive_frequency_1st_sum_rule} shows that 
the first moment selects the momentum-odd part of
\(C^{xy}(\bq,\omega)\) and is fixed by a momentum-weighted static vector
chirality.
The product
\(\sin[\bq\cdot(\br_i-\br_j)]
(\mathbf S_i\times\mathbf S_j)^z\)
is invariant under \(i\leftrightarrow j\), so the result is
independent of the arbitrary orientation assigned to each bond.
The transverse exchange contributions proportional to
\(J^\perp_{ij}\) cancel between the two double commutators, so only the
longitudinal exchange \(J^z_{ij}\) contributes.
Consequently, the first moment vanishes in the pure-XY limit
\(J^z_{ij}=0\), even when the state has finite vector chirality.

\section{Collinear antiferromagnets}

We apply the general framework to collinear antiferromagnets with zero
uniform magnetization.
We consider models with Heisenberg or XXZ-type exchange but without
Dzyaloshinskii--Moriya interactions or other terms that break the
residual spin-only symmetries considered below.
We choose the collinear order parameter along the \(z\) axis, so 
the ordered state retains a U(1) spin-rotation symmetry about this axis.

In this setting, moment constraints must be distinguished 
from spectral selection rules.
Vanishing frequency moments do not imply that the antisymmetric DSF
itself vanishes, because each moment probes only a particular
momentum-parity combination of the spectrum.
We therefore first determine the allowed tensor components and their
momentum parity from the residual SSG and then combine
these results with the moment relations and sum rules derived in
Sec.~\ref{sec:sum_rules}.

\subsection{Symmetry constraints and classification}

Chen {\it et al.} organized collinear spin-space groups (SSGs) into four types and
used SSG band representations and little-group co-representations to
classify symmetry-enforced magnon-band degeneracies and symmetry-allowed chirality splitting~\cite{Chen2025SSGMagnons}.
Type I describes collinear ferromagnets and ferrimagnets and lies outside the present setting 
of compensated antiferromagnets with zero uniform magnetization.
We therefore focus on types II--IV, for which a spatial operation $A$ relates the two opposite-spin sublattices.
This relation can be represented schematically by
\begin{equation}
(C^s_{2\mathbf{n}}\Vert A),
\end{equation}
where \(C^s_{2\mathbf n}\) is a spin \(\pi\) rotation about an axis
\(\mathbf n\perp\hat{\mathbf z}\) and therefore reverses the ordered moment.
The combined operation leaves the collinear spin structure invariant.

The type is determined by the spatial part \(A\).  In the notation of
Ref.~\cite{Chen2025SSGMagnons}, \(A=P\) gives type II, 
\(A=T_{\boldsymbol\tau}\) gives type IV, and 
\(A=C_m\) or \(PC_m\), with \(m=2,4\), gives type III.
Here \(P\) denotes spatial inversion, \(T_{\boldsymbol\tau}\) a translation, and
\(C_m\) an \(m\)-fold spatial rotation.
Type III corresponds to the altermagnetic class. 
We use this SSG classification to determine whether the spin-component-antisymmetric 
DSF is forbidden or allowed by symmetry.

The selection rules for \(C^{\alpha\beta}(\bq,\omega)\) are determined
by the full residual SSG.
Because the ordered moment is chosen along the \(z\) axis, the residual SSG
contains the spin-only unitary operation \((C^s_{2z}\Vert E)\).  
Equation~\eqref{eq:unified_unitary_C_constraint} then gives
\begin{equation}
C^{xz}(\mathbf q,\omega)
=
C^{yz}(\mathbf q,\omega)
=
0 .
\end{equation}
Thus \(C^{xy}\) is the only antisymmetric component that can
remain nonzero.

The collinear state also has a spin-only antiunitary symmetry, for
example \((\theta C_{2x}^s\Vert E)\), for which the spin \(\pi\) rotation compensates
the spin reversal under time reversal.
For the remaining \(xy\) component,
Eq.~\eqref{eq:unified_antiunitary_C_constraint} gives
\begin{equation}
C^{xy}(\bq,\omega)=C^{xy}(-\bq,\omega).
\end{equation}
Thus, \(C^{xy}\) is even under momentum reversal.
Combining this parity with Eq.~\eqref{eq:C_positive_frequency_moments} gives
\(M_1^{xy}(\bq)=0\).
For the other antisymmetric components, the stronger spectral selection rule
\(C^{xz}=C^{yz}=0\) directly gives \(M_1^{xz}=M_1^{yz}=0\). Hence,
\[
M_1^{\alpha\beta}(\bq)=0
\]
for all antisymmetric components.

Equivalently, the same antiunitary symmetry makes
\((\mathbf S_i\times\mathbf S_j)^z\) odd and therefore forces its
expectation value to vanish.
The vector-chirality form of the first-moment sum rule,
Eq.~\eqref{eq:positive_frequency_1st_sum_rule}, then gives the same
result.

By contrast, the vanishing zeroth moment follows directly from the
zeroth-moment sum rule and the absence of uniform magnetization:
\[
M_0^{\alpha\beta}(\bq)=0.
\]
For the only potentially nonzero component, combining
$M_0^{xy}(\bq)=0$ with the symmetry-enforced evenness of \(C^{xy}\)
reduces the zeroth-moment relation, at each $\bq$, to
\begin{equation}\label{eq:altmag_zero_moment_relration}
\int_0^\infty d\omega\,
C^{xy}(\bq,\omega)=0.
\end{equation}

The type-dependent selection rules follow from the sublattice-exchanging operation
\((C^s_{2\mathbf n}\Vert A)\).
Combining this operation with the spin-only antiunitary symmetry
\((\theta C^s_{2\mathbf n}\Vert E)\) generates \((\theta\Vert A)\).
This antiunitary operation determines whether the remaining momentum-even
component \(C^{xy}\) is forbidden throughout the Brillouin zone or can remain finite 
at generic momenta with symmetry-enforced nodes.

\paragraph{Type II.}

For type-II SSGs, \(A=P\), so the residual SSG contains the
antiunitary inversion symmetry \((\theta\Vert P)\).
As discussed in Sec.~\ref{sec:corol}, this symmetry forces all off-diagonal components of
the antisymmetric DSF to vanish:
\begin{equation}
C^{\alpha\beta}(\mathbf q,\omega)=0,
\qquad
\alpha\ne\beta .
\end{equation}

\paragraph{Type IV.}

For type-IV SSGs, \(A=T_{\boldsymbol\tau}\) is a pure translation.
Choosing the \(x\) axis along \(\mathbf n\), we write the
sublattice-exchanging operation as
\((C_{2x}^s\Vert T_{\boldsymbol\tau})\).
Multiplying it by the spin-only symmetry \((C_{2z}^s\Vert E)\) generates
\((C_{2y}^s\Vert T_{\boldsymbol\tau})\).
As discussed in Sec.~\ref{sec:corol}, these two orthogonal
spin-\(\pi\)-rotation symmetries forbid all off-diagonal components of the
antisymmetric DSF:
\begin{equation}
C^{\alpha\beta}(\bq,\omega)=0, \qquad \alpha \ne \beta.
\end{equation}

\paragraph{Type III.}

For type-III SSGs, \(A=C_m\) or \(PC_m\), with \(m=2,4\), so the
residual SSG contains \((\theta\Vert A)\).
For the remaining \(xy\) component, Eq.~\eqref{eq:unified_antiunitary_C_constraint} gives
\[
C^{xy}(\bq,\omega)
=
-
C^{xy}(-R_A\bq,\omega).
\]
Together with the evenness under momentum reversal,
\(C^{xy}(\bq,\omega)=C^{xy}(-\bq,\omega)\), this relation becomes
\begin{equation}
C^{xy}(\bq,\omega)
=
-
C^{xy}(R_A\bq,\omega).
\end{equation}
Thus, \(C^{xy}\) can be finite at generic momenta in type-III
collinear antiferromagnets but has symmetry-enforced nodes whenever
\[
R_A\bq=\pm\bq.
\]
More generally, an additional antiunitary operation
\((\theta\Vert B)\) in the residual SSG imposes the same type of
constraint, with \(R_A\) replaced by \(R_B\).

\begin{table}[t]
\centering
\renewcommand{\arraystretch}{1.3}
\caption{Classification of collinear antiferromagnets by the SSG
operation \((C_{2\mathbf n}^s\Vert A)\), where \(A\) exchanges
opposite-spin sublattices.
The table shows the resulting constraints on the
spin-component-antisymmetric DSF.}
\label{tab:ssg_classification_C}

\begin{tabular*}{\linewidth}{@{\extracolsep{\fill}}c c l}
\hline\hline
SSG class &
\(A\) &
constraint on \(C^{\alpha\beta}\) \((\alpha \ne \beta)\)
\\
\hline

type II &
\(P\) &
\( C^{\alpha\beta}(\bq,\omega)=0\)
\\[5pt]

type IV &
\(T_{\boldsymbol\tau}\) &
\( C^{\alpha\beta}(\bq,\omega)=0\)
\\[5pt]

type III &
\begin{tabular}{@{}l@{}}
\(C_m\) or \(P C_m\), \\
\(m=2,4\)
\end{tabular} &
\begin{tabular}{@{}l@{}}
\(C^{xy}\) can be nonzero at generic \(\bq\); \\
even under \(\bq\to-\bq\), with nodes \\
where \(R_A\bq=\pm\bq\)
\end{tabular}
\\

\hline\hline
\end{tabular*}
\end{table}

These results are summarized in Table~\ref{tab:ssg_classification_C}.
Here we complement the magnon-band analysis of Ref.~\cite{Chen2025SSGMagnons}
by applying the type-II--IV classification directly to the dynamical spin-correlation
tensor.  
The resulting vanishing rules, momentum-parity constraints,
and symmetry-enforced nodes therefore do not require well-defined
magnon quasiparticles; they follow whenever the Hamiltonian and the
selected ordered state possess the assumed residual SSG symmetries.

\subsection{Type-III altermagnet}
\label{sec:altermag}

Previous theoretical work on polarized-neutron scattering showed that, in an ideal
two-sublattice altermagnet with a global U(1) spin-rotation symmetry,
altermagnetic exchange splits magnon branches with opposite circular
polarizations~\cite{McClarty2025}.
The two branches have equal unpolarized intensities but opposite
polarization-dependent weights.
Related chiral magnon splitting has been observed in
MnTe~\cite{Liu2024MnTe,Liu2026MnTe}.
In the present notation, this branch-resolved mechanism produces a
finite \(C^{xy}(\bq,\omega)\) at generic momenta.
We use a minimal model to connect this mechanism explicitly to the residual SSG
constraints and the exact sum rules for the commutator moments derived above.
The analysis identifies the symmetry-enforced nodes, the branch cancellation that
realizes the vanishing zeroth moment, and the even parity under momentum reversal
that enforces the vanishing first moment.

\subsubsection{Model}

\begin{figure}
  \centering
  \includegraphics[width=7cm]{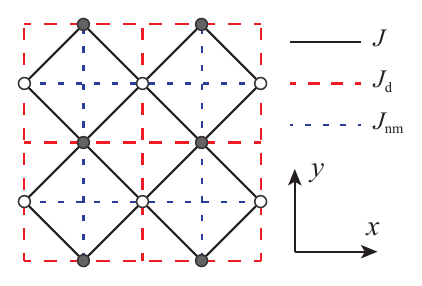}
  \caption{Minimal two-sublattice square-lattice model used in the
linear spin-wave calculation.
Solid, red dashed, and blue dotted bonds denote
\(J\), \(J_{\rm d}\), and \(J_{\rm nm}\), respectively.
Gray and open circles denote the $A$ and $B$ sublattices, respectively.
The primitive lattice vectors are
\(\mathbf a_1=2a\,\hat{\mathbf x}\) and
\(\mathbf a_2=2a\,\hat{\mathbf y}\).}\label{fig:model}
\end{figure}

We consider the effective localized-spin model for the two magnetic
sublattices of the square-lattice altermagnet introduced in
Ref.~\cite{BrekkeBrataasSudbo2023}.
We choose the Cartesian axes shown in Fig.~\ref{fig:model}.
The primitive lattice vectors are
\(\mathbf a_1=2a\,\hat{\mathbf x}\) and
\(\mathbf a_2=2a\,\hat{\mathbf y}\),
where \(a\) sets the length scale. We set $a=1$ below.
The exchange Hamiltonian is
\begin{align}
H
&=
J\sum_{\langle i,j\rangle}
\mathbf S_i\cdot\mathbf S_j
+
J_{\rm d}\sum_{\langle\!\langle i,j\rangle\!\rangle_{\rm d}}
\mathbf S_i\cdot\mathbf S_j
+
J_{\rm nm}\sum_{\langle\!\langle i,j\rangle\!\rangle_{\rm nm}}
\mathbf S_i\cdot\mathbf S_j,
\label{eq:altmag_model}
\end{align}
where \(J>0\) and \(J_{\rm d},J_{\rm nm}\le 0\).
The coupling \(J\) connects the two opposite-spin sublattices,
whereas \(J_{\rm d}\) and \(J_{\rm nm}\) are the two inequivalent
same-sublattice couplings shown in Fig.~\ref{fig:model}.
Following
Ref.~\cite{BrekkeBrataasSudbo2023}, \(J_{\rm nm}\) denotes exchange
mediated by an intermediate nonmagnetic site, whereas \(J_{\rm d}\) denotes
direct exchange between magnetic sites without such an intermediate
site.
The localized-spin Hamiltonian of
Ref.~\cite{BrekkeBrataasSudbo2023} also contains a Zeeman field
\(B_z\) and an easy-axis anisotropy \(K_z\).
We set both terms to zero and focus on the SU(2)-symmetric exchange
model.

We take the ordered moment along the \(z\) axis.  By the general
collinear constraints derived above, \(C^{xy}\) is the only component that can remain nonzero.
We therefore focus on this component and examine the additional constraints imposed by 
the model-specific SSG operations.

In addition to the spin-only symmetries used above,
the residual SSG of the ordered state contains the following model-specific operations:
\begin{align}
&(E_s\Vert C_{2z}),\quad
(E_s\Vert \sigma_x),\quad
(E_s\Vert \sigma_y),\nonumber\\
&(\theta\Vert C_{4z}^{\pm}),\quad
(\theta\Vert \sigma_{xy}),\quad
(\theta\Vert \sigma_{x\bar y}).
\label{eq:altmag_ssg_operations}
\end{align}
Here \(E_s\) denotes the identity operation in spin space.
The symbols \(\sigma_x\) and \(\sigma_y\) denote reflections across the
\(x\) and \(y\) axes, respectively, whereas \(\sigma_{xy}\) and
\(\sigma_{x\bar y}\) denote reflections across the two diagonal lines.
These operations impose the model-specific momentum constraints
on \(C^{xy}(\bq,\omega)\) derived below.

\subsubsection{Symmetry analysis}

We first consider the unitary mirror symmetries.  
Equation~\eqref{eq:unified_unitary_C_constraint} shows that
\((E_s\Vert\sigma_x)\) and \((E_s\Vert\sigma_y)\) give
\begin{align}
(E_s\Vert \sigma_x):\quad
C^{xy}(q_x,q_y,\omega)
&=
C^{xy}(q_x,-q_y,\omega),
\label{eq:altmag_ssg_sigmax}
\\
(E_s\Vert \sigma_y):\quad
C^{xy}(q_x,q_y,\omega)
&=
C^{xy}(-q_x,q_y,\omega).
\label{eq:altmag_ssg_sigmay}
\end{align}
Thus, \(C^{xy}\) is even under independent sign changes of \(q_x\) and
\(q_y\).

We next consider the antiunitary diagonal mirrors
\((\theta\Vert\sigma_{xy})\) and
\((\theta\Vert\sigma_{x\bar y})\).
In the notation of the general type-III analysis, these
correspond to the additional antiunitary operations
\((\theta\Vert B)\), with
\(B=\sigma_{xy}\) and \(B=\sigma_{x\bar y}\), respectively.
The two diagonal reflections act as
\[
R_{\sigma_{xy}}(q_x,q_y)=(q_y,q_x),
\qquad
R_{\sigma_{x\bar y}}(q_x,q_y)=(-q_y,-q_x).
\]
Combining Eq.~\eqref{eq:unified_antiunitary_C_constraint} with the
evenness under momentum reversal,
\(C^{xy}(\bq,\omega)=C^{xy}(-\bq,\omega)\), both symmetries impose
the same constraint:
\begin{equation}
C^{xy}(q_x,q_y,\omega)
=
-C^{xy}(q_y,q_x,\omega).
\label{eq:altmag_ssg_exchange_odd}
\end{equation}
Thus, \(C^{xy}\) is odd under interchange of \(q_x\) and \(q_y\).

For each diagonal reflection, the condition
\(R_B\bq=\pm\bq\) is satisfied along the two diagonals \(q_x=\pm q_y\).  
The antiunitary symmetries therefore enforce the line nodes
\begin{equation}
C^{xy}(\bq,\omega)=0
\qquad\text{for}\qquad
q_x=\pm q_y.
\end{equation}
Thus, the model-specific SSG operations impose a
\(d_{x^2-y^2}\)-type momentum structure with line nodes at
\(q_x=\pm q_y\), while allowing a finite \(C^{xy}\) at generic momenta.

\subsubsection{Linear spin-wave analysis}

We next evaluate \(C^{xy}\) within linear spin-wave theory (LSWT).
For the SU(2)-symmetric exchange model in Eq.~\eqref{eq:altmag_model}, the
quadratic spin-wave Hamiltonian gives two transverse magnon branches:
\begin{equation}
\omega_{\alpha,\bq}
=
\Omega_{\bq}+\gamma_{1,\bq},
\qquad
\omega_{\beta,\bq}
=
\Omega_{\bq}-\gamma_{1,\bq},
\end{equation}
where
\[
\Omega_{\bq}
=
\sqrt{\gamma_{2,\bq}^2-\gamma_{3,\bq}^2}.
\]
The coefficients \(\gamma_{1,\bq}\), \(\gamma_{2,\bq}\), and
\(\gamma_{3,\bq}\) are defined in Appendix~\ref{sec:LSWT_altmag_append}.
The splitting between the two branches is
\[
\omega_{\alpha,\bq}-\omega_{\beta,\bq}
=
2\gamma_{1,\bq}.
\]
Because
$\gamma_{1,\bq}\propto \cos(2q_x)-\cos(2q_y)$,
the splitting has
\(d_{x^2-y^2}\)-type symmetry and vanishes along the Brillouin-zone diagonals.

The dimensionless one-magnon form factor for the transverse spectral weight is
\begin{equation}
F_{\mathbf q}
=
\sqrt{
\frac{\gamma_{2,\mathbf q}-\gamma_{3,\mathbf q}}
{\gamma_{2,\mathbf q}+\gamma_{3,\mathbf q}}
}.
\label{eq:altmag_form_factor}
\end{equation}
The two transverse branches carry opposite circular polarizations and
therefore contribute with opposite signs to the antisymmetric DSF:
\begin{equation}
C^{xy}_{1{\rm mag}}(\mathbf q,\omega)
=
\frac{S}{4}
F_{\mathbf q}
\left[
\delta(\omega-\omega_{\alpha,\mathbf q})
-
\delta(\omega-\omega_{\beta,\mathbf q})
\right].
\label{eq:altmag_Cxy}
\end{equation}
The matrix-element derivation is given in Appendix~\ref{sec:LSWT_altmag_append}.
When the two inequivalent same-sublattice exchanges are equal,
\(J_{\rm d}=J_{\rm nm}\), the oppositely polarized branches are degenerate
and their contributions to \(C^{xy}\) cancel.
For \(J_{\rm d}\ne J_{\rm nm}\), their \(d_{x^2-y^2}\)-type splitting separates
them in energy and produces a finite \(C^{xy}\) at generic momenta.
Equation~\eqref{eq:altmag_Cxy} explicitly realizes this momentum
structure: \(C^{xy}_{1{\rm mag}}\) changes sign under
\(q_x\leftrightarrow q_y\) and vanishes on \(q_x=\pm q_y\).

The one-magnon form factor vanishes linearly near the $\Gamma$ point,
$F_{\bq}
\propto |\bq|$.
This behavior follows because
\(\gamma_{2,\bq}-\gamma_{3,\bq}\) vanishes quadratically in \(\bq\),
whereas \(\gamma_{2,\bq}+\gamma_{3,\bq}\) remains finite.
Consequently, the magnitude of the antisymmetric weight carried by each circularly polarized branch
vanishes linearly as \(|\bq|\to0\).

Although the split branches produce a nonzero frequency-resolved
\(C^{xy}\), Eq.~\eqref{eq:altmag_Cxy} shows that they carry equal and
opposite integrated weights:
\begin{equation}
\int_0^\infty d\omega\,
C^{xy}_{1{\rm mag}}(\bq,\omega)=0 .
\end{equation}
Within LSWT, this branch cancellation explicitly
realizes the integral constraint in
Eq.~\eqref{eq:altmag_zero_moment_relration}, which follows from the vanishing
zeroth moment and the even parity of \(C^{xy}\) under momentum reversal.
Independently of spin-wave theory, this even parity, together with
Eq.~\eqref{eq:C_positive_frequency_moments}, gives
\(M_1^{xy}(\bq)=0\).

\section{Planar helical magnet}

The type-III altermagnet discussed in the previous section provides an
example in which \(C^{xy}\) is finite even though
the zeroth and first moments of the corresponding commutator spectral function
vanish.  We now contrast this behavior with that of an ideal planar
helical state,
\begin{equation}
\left\langle \mathbf S_{\mathbf r} \right\rangle
=
S\bigl(
\cos(\mathbf Q\cdot\mathbf r),\sin(\mathbf Q\cdot\mathbf r),0
\bigr),
\label{eq:ideal_planar_helix}
\end{equation}
in a time-reversal-invariant exchange Hamiltonian with a U(1) spin-rotation symmetry about the $z$ axis.
We assume a genuinely noncollinear helix, excluding both the ferromagnetic 
and two-sublattice antiferromagnetic limits.

The SSG analysis below determines the allowed tensor components and 
their parity under momentum reversal.
We then combine these spectral constraints with the exact moment relations 
and sum rules.
The zeroth-moment sum rule gives
\(M_0^{xy}(\bq)=0\) because the uniform magnetization vanishes.
By contrast, \(M_1^{xy}(\bq)\) is allowed by symmetry and, for Heisenberg
or XXZ exchange Hamiltonians, is fixed by a momentum-weighted static vector
chirality through the first-moment sum rule.

\subsection{SSG constraints and moment implications}
\label{sec:SSG_helical}
Let \(\mathcal L\) be the Bravais lattice.
With the translation convention
$T_{\mathbf t}\mathbf S_{\mathbf r}T_{\mathbf t}^{-1}
=
\mathbf S_{\mathbf r+\mathbf t}$,
the helical texture in Eq.~\eqref{eq:ideal_planar_helix} is invariant under
the combined translation--spin-rotation operations
\begin{equation}
\left\{
(R_z^s(-\mathbf Q\cdot\mathbf t)\Vert T_{\mathbf t})
\,\middle|\,
\mathbf t\in \mathcal L
\right\},
\label{eq:helix_ssg_translation}
\end{equation}
where \(R_z^s(\phi)\) denotes a spin rotation about the
\(z\) axis by an angle \(\phi\).

We first apply the unitary constraints associated with
Eq.~\eqref{eq:helix_ssg_translation}.
Because the real-space part of each operation is a pure translation,
it does not change the momentum \(\bq\) in the constraint on
\(C^{\alpha\beta}(\bq,\omega)\).
Applying Eq.~\eqref{eq:unified_unitary_C_constraint} to
\((R_z^s(\phi_{\mathbf t})\Vert T_{\mathbf t})\), with
\(\phi_{\mathbf t}=-\mathbf Q\cdot\mathbf t\), gives
\begin{equation}
C^{\alpha\beta}(\bq,\omega)
=
R^z_{\alpha\gamma}(\phi_{\mathbf t})
R^z_{\beta\delta}(\phi_{\mathbf t})
C^{\gamma\delta}(\bq,\omega),
\label{eq:helix_unitary_constraint_tensor}
\end{equation}
where \(R^z(\phi_{\mathbf t})\) is the spin-space matrix representing
\(R_z^s(\phi_{\mathbf t})\).
For the antisymmetric tensor, this condition is equivalent to invariance of
the corresponding axial vector
\(C^\mu_{\rm ax}=\epsilon_{\mu\alpha\beta}C^{\alpha\beta}/2\) under
\(R_z(\phi_{\mathbf t})\).
For a genuinely noncollinear helix, there is a translation
\(\mathbf t\) for which
\(\phi_{\mathbf t}\not\equiv0,\pi \pmod{2\pi}\).
Invariance under this spin rotation allows only the axial-vector
component parallel to the helical axis.
Consequently,
\begin{equation}
C^{xz}(\mathbf q,\omega)=C^{yz}(\mathbf q,\omega)=0,
\end{equation}
and only \(C^{xy}\) can remain finite.

The planar helical texture is also invariant under the spin-only
antiunitary symmetry
\((\theta R_z^s(\pi)\Vert E)\), because the spin \(\pi\) rotation
restores the planar spin texture after time reversal.
The rotation \(R_z^s(\pi)\) changes the signs of both the \(x\) and \(y\)
spin components, so its tensor action leaves \(C^{xy}\) invariant.
Equation~\eqref{eq:unified_antiunitary_C_constraint} therefore gives
\begin{equation}
C^{xy}(\bq,\omega)
=
-
C^{xy}(-\bq,\omega).
\label{eq:helix_Cxy_odd_q}
\end{equation}
Thus, in the ideal planar helix, the remaining \(xy\) component of the
antisymmetric DSF is odd under momentum reversal as a consequence of this
residual combined antiunitary symmetry.

Combining this odd parity with Eq.~\eqref{eq:C_positive_frequency_moments}
gives
\[
M_0^{xy}(\bq)=0,
\]
because the zeroth moment probes the momentum-even part of
\(C^{xy}\).
The same result follows independently from the zeroth-moment sum rule
because the uniform magnetization along the helical axis vanishes.

By contrast, the odd parity does not force the first moment to vanish,
because the first moment probes the momentum-odd part of \(C^{xy}\).
For Heisenberg or XXZ exchange Hamiltonians, the first-moment sum rule relates
\(M_1^{xy}(\bq)\) to a momentum-weighted static vector chirality.
The bond vector chirality
\(\langle(\mathbf S_i\times\mathbf S_j)^z\rangle\) is allowed by symmetry and is
generally finite in a single-domain planar helix.
Consequently, \(M_1^{xy}(\bq)\) is generally nonzero at generic
momenta.

\subsection{Triangular-lattice helical magnet}
\label{sec:LSWT_helical}

As a concrete realization of an ideal planar helix in an
SU(2)-symmetric exchange model, we consider the spatially anisotropic
triangular-lattice Heisenberg antiferromagnet~\cite{Starykh2015}.
The spin-component-antisymmetric response of triangular antiferromagnets
has previously been studied in the context of dynamical chirality in
polarized-neutron scattering.  In particular, Syromyatnikov derived the
linear spin-wave expression for the dynamical chirality of triangular
antiferromagnets and showed that this response can remain nonzero in a single chiral
domain even without Dzyaloshinskii--Moriya interactions or an applied magnetic
field~\cite{Syromyatnikov2005}.
Related chiral spin-wave intensities have been measured in
Ba$_3$NbFe$_3$Si$_2$O$_{14}$~\cite{Loire2011}.

We use the standard rotating-frame linear spin-wave theory (LSWT) for this
model and express its
one-magnon inelastic response in terms of the antisymmetric DSF
\(C^{xy}(\bq,\omega)\).
This formulation connects the earlier LSWT result for
dynamical chirality with both the odd parity under momentum reversal derived here
from the residual SSG and the moment relations and sum
rules derived in Sec.~\ref{sec:sum_rules}.

The Hamiltonian is
\begin{equation}
H
=
\sum_{\mathbf r}
\left[
J\,\mathbf S_{\mathbf r}\cdot\mathbf S_{\mathbf r+\boldsymbol\delta_1}
+
J'\,\mathbf S_{\mathbf r}\cdot\mathbf S_{\mathbf r+\boldsymbol\delta_2}
+
J'\,\mathbf S_{\mathbf r}\cdot\mathbf S_{\mathbf r+\boldsymbol\delta_3}
\right],
\label{eq:H_helical}
\end{equation}
where \(J,J'>0\),
$\boldsymbol\delta_1=(1,0)$,
$\boldsymbol\delta_2=(1/2,\sqrt{3}/2)$, and
$\boldsymbol\delta_3=(1/2,-\sqrt{3}/2)$.
For a classical planar spiral with wave vector \(\mathbf Q\), the energy
per site is \(S^2 J_{\mathbf Q}\), where \(J_{\mathbf q}\) is the Fourier
transform of the exchange couplings:
\begin{equation}\label{eq:Jq_triangular}
J_{\bq}
=
J\cos q_x
+
2J'\cos\frac{q_x}{2}\cos\frac{\sqrt3 q_y}{2}.
\end{equation}
The classical ordering wave vector minimizes
\(J_{\bq}\).  For \(0<J'< 2J\), it can be chosen as
\begin{equation}
\mathbf Q=(Q,0),
\qquad
\cos\frac{Q}{2}=-\frac{J'}{2J}.
\end{equation}
The isotropic triangular-lattice limit \(J'=J\) gives \(Q=4\pi/3\), corresponding
to the \(120^\circ\) state.

The standard rotating-frame LSWT gives the magnon dispersion
\begin{equation}
\omega_{\bq}
=
S
\sqrt{
2
\left(J_{\bq}-J_{\mathbf Q}\right)
\left(
J_{\bq+\mathbf Q}
+
J_{\bq-\mathbf Q}
-
2J_{\mathbf Q}
\right)
}.
\end{equation}
The transverse one-magnon spectral weight is
\begin{equation}
W_{\bk}
=
S^2
\frac{
J_{\bk}-J_{\mathbf Q}
}{
\omega_{\bk}
}.
\label{eq:helix_Wk}
\end{equation}
Because the physical transverse spin operators at momentum \(\bq\)
couple to rotating-frame magnons at \(\bq\pm\mathbf Q\), the
one-magnon contribution to the antisymmetric DSF is
\begin{align}
C^{xy}_{1{\rm mag}}(\mathbf q,\omega)
=
\frac{1}{4}
[
&W_{\mathbf q+\mathbf Q}\delta(\omega-\omega_{\mathbf q+\mathbf Q})\nonumber\\
& - W_{\mathbf q-\mathbf Q}\delta(\omega-\omega_{\mathbf q-\mathbf Q})
].
\label{eq:helix_Cxy_LSWT}
\end{align}
The matrix-element derivation is given in Appendix~\ref{sec:LSWT_helical_append}.

The singular behavior of Eq.~\eqref{eq:helix_Cxy_LSWT} is controlled by the
spectral weight \(W_{\mathbf k}\).
Near \(\bk=0\),
\(J_{\bk}-J_{\mathbf Q}\) remains finite for a genuine noncollinear
helix, whereas \(\omega_{\bk}=O(|\bk|)\).  Hence,
\(W_{\bk}\propto 1/|\bk|\).
In Eq.~\eqref{eq:helix_Cxy_LSWT}, this singularity is reached when either
\(\bq-\bQ\) or \(\bq+\bQ\) approaches zero.
Thus, the one-magnon spectral weights in \(C^{xy}_{1{\rm mag}}\) diverge with opposite signs
near \(\bq=\mathbf Q\) and \(\bq=-\mathbf Q\).
This Goldstone-mode enhancement is qualitatively analogous to the phase-mode singularity discussed for
itinerant helimagnets in Ref.~\cite{BelitzKirkpatrickRosch2006}.

By contrast, near the physical zone center \(\bq=0\), the shifted
momenta \(\bq\pm\mathbf Q\) approach the gapless points
\(\pm\mathbf Q\).  Because
\(J_{\mathbf Q\pm\bq}-J_{\mathbf Q}=O(q^2)\) and
\(\omega_{\mathbf Q\pm\bq}=O(|\bq|)\), one finds
\(W_{\mathbf Q\pm\bq}\propto |\bq|\).
Thus, the corresponding one-magnon spectral weights
vanish linearly near \(\bq=0\).

Equation~\eqref{eq:helix_Cxy_LSWT} explicitly realizes the odd parity under momentum reversal 
derived in Sec.~\ref{sec:SSG_helical} for a fixed helical
domain:
\[
C^{xy}_{1{\rm mag}}(\bq,\omega)
=
-
C^{xy}_{1{\rm mag}}(-\bq,\omega).
\]
The opposite-helicity domain, obtained by \(\mathbf Q\to-\mathbf Q\), has
the opposite sign of \(C^{xy}\) at fixed \(\bq\).

We finally compare the LSWT result with the exact moment relations and sum rules derived in Sec.~\ref{sec:sum_rules}.
The commutator zeroth moment vanishes as a direct consequence of the
odd parity under momentum reversal, so we focus on the first moment.
Equation~\eqref{eq:C_positive_frequency_moments} gives the leading-order
spin-wave contribution
\begin{align}
M^{xy}_{1,\mathrm{LSWT}}(\mathbf q)
&=
2\int_0^\infty d\omega\,
\omega C^{xy}_{1{\rm mag}}(\mathbf q,\omega)
\nonumber\\
&=
\frac{S^2}{2}
\left(
J_{\mathbf q+\mathbf Q}
-
J_{\mathbf q-\mathbf Q}
\right)
\nonumber\\
&=
-S^2\sum_{\boldsymbol\delta}
J_{\boldsymbol\delta}
\sin(\mathbf q\cdot\boldsymbol\delta)
\sin(\mathbf Q\cdot\boldsymbol\delta).
\label{eq:helix_first_moment_LSWT}
\end{align}
Here the sum runs over the three bond vectors
\(\boldsymbol\delta_1,\boldsymbol\delta_2,\boldsymbol\delta_3\)
defined above.
In the classical helical state,
\begin{equation}
\left\langle
(\mathbf S_{\mathbf r}\times\mathbf S_{\mathbf r+\boldsymbol\delta})^z
\right\rangle
=
S^2\sin(\mathbf Q\cdot\boldsymbol\delta).
\end{equation}
Substituting this classical vector chirality into the exact sum rule,
Eq.~\eqref{eq:1st_sum_rule}, reproduces
Eq.~\eqref{eq:helix_first_moment_LSWT}.
Thus, the LSWT result realizes the vector-chirality sum rule at
leading order in \(S\).

The planar helix therefore contrasts with the type-III
altermagnet: \(C^{xy}\) is odd under momentum reversal and has a generally
nonzero first moment fixed by a momentum-weighted static
vector chirality.

\section{Summary}

We have studied the spin-component-antisymmetric part of the dynamical 
spin-structure-factor tensor, \(C^{\alpha\beta}(\mathbf q,\omega)\), focusing on its
symmetry constraints and frequency-moment sum rules.
We derived general constraints imposed by unitary and antiunitary operations of magnetic space groups
and spin-space groups.
These constraints determine the allowed tensor components, their parity under momentum reversal, 
and symmetry-enforced nodes.  
We also introduced the corresponding antisymmetric commutator spectral
function and derived exact sum rules for its zeroth and first moments. 
The zeroth moment is fixed by the uniform magnetization.
For Heisenberg or XXZ exchange models, the first moment of the $xy$ component
is fixed by a momentum-weighted static vector chirality.

Applying these results to collinear antiferromagnets with the ordered moment along the $z$ axis,
we found that the spin-only symmetries allow at most the \(xy\) component.
Within the classification of Chen {\it et al.}, this remaining component
is also forbidden by symmetry in types II and IV.
In type III, the altermagnetic class, \(C^{xy}(\mathbf q,\omega)\) is allowed at generic
momenta but has symmetry-enforced nodes.
In the square-lattice altermagnetic example, a finite \(C^{xy}(\mathbf q,\omega)\) arises from split
magnon branches with opposite circular polarizations.
The cancellation between these branches realizes the vanishing zeroth moment, 
whereas the first moment vanishes because \(C^{xy}\) is even under momentum reversal.

We also analyzed a planar helical magnet as a contrasting example.
A residual spin-only antiunitary symmetry makes
\(C^{xy}(\bq,\omega)\) odd under momentum reversal.
This odd parity forces the commutator zeroth moment to vanish, consistent
with the absence of uniform magnetization.
By contrast, the first moment
is generally nonzero and is fixed by a momentum-weighted static vector chirality of the
helical spin texture.
In the triangular-lattice example, the one-magnon spin-wave response explicitly
realizes these constraints and satisfies the vector-chirality sum rule
at leading order in the spin-wave expansion.

\begin{table}[t]
\centering
\renewcommand{\arraystretch}{1.25}
\caption{
Representative ordered states for which \(C^{xy}(\mathbf q,\omega)\) can be finite.
The table lists the microscopic origin
of the signal, its parity under momentum reversal \(\mathbf q\to-\mathbf q\),
and the zeroth and first moments of the corresponding antisymmetric
commutator spectral function \(\mathcal A^{xy}_{\rm as}\).
The ordered moment in the collinear state and the helical axis are taken along the \(z\) axis.
A collinear ferromagnet (FM) with inversion symmetry is
included for comparison.
}
\label{tab:mechanisms_Cxy}
\begin{tabular*}{\columnwidth}{@{\extracolsep{\fill}}lcccc}
\hline\hline
state & origin & parity & \(M_0^{xy}\) & \(M_1^{xy}\) \\
\hline
Collinear FM & magnetization & even & \(\neq 0\) & \(0\) \\
Altermagnet & magnon-band splitting & even & \(0\) & \(0\) \\
Planar helix & vector chirality & odd & \(0\) & \(\neq 0\) \\
\hline\hline
\end{tabular*}
\end{table}

Table~\ref{tab:mechanisms_Cxy} summarizes three distinct mechanisms
for producing a finite antisymmetric DSF: uniform magnetization,
magnon-band splitting, and static vector chirality.
The momentum parity and the zeroth and first moments associated with
each mechanism provide a unified way to distinguish the microscopic
origins of chiral dynamical spin correlations.

\acknowledgments
The author thanks Kazushi Aoyama and Hitoshi Seo for stimulating discussions.

\appendix

\section{Linear spin-wave calculation for the two-dimensional altermagnetic model}
\label{sec:LSWT_altmag_append}

We derive the one-magnon contribution to the
antisymmetric DSF for the
two-dimensional altermagnetic model discussed in Sec.~\ref{sec:altermag}.
We choose the ordered moment along the \(z\) axis.

For the two sublattices \(A\) and \(B\), we use the Holstein--Primakoff
transformation to the order required for linear spin-wave theory (LSWT):
\begin{align}
S_i^+
&=
\sqrt{2S}\,a_i,
&
S_i^-
&=
\sqrt{2S}\,a_i^\dagger,
&
S_i^z
&=
S-a_i^\dagger a_i,
\quad i\in A,
\nonumber\\
S_j^+
&=
\sqrt{2S}\,b_j^\dagger,
&
S_j^-
&=
\sqrt{2S}\,b_j,
&
S_j^z
&=
-S+b_j^\dagger b_j,
\quad j\in B .
\end{align}
Let \(N\) be the number of magnetic unit cells.  For the boson operators, we use
\begin{equation}
a_{\bq}
=
\frac{1}{\sqrt N}
\sum_{\ell\in A} e^{-i\bq\cdot \mathbf r_\ell}a_\ell,
\quad
b_{\bq}
=
\frac{1}{\sqrt N}
\sum_{\ell \in B} e^{-i\bq\cdot \mathbf r_\ell}b_\ell .
\end{equation}
For the spin operators, we use
\begin{equation}
S^\alpha_{\bq}
=
\frac{1}{\sqrt{2 N}}
\sum_{\ell\in A\cup B}
e^{-i\bq\cdot \mathbf r_\ell}
S^\alpha_\ell .
\end{equation}
With these conventions,
\((S^+_{\bq})^\dagger=S^-_{-\bq}\).
To leading order, the transverse spin operators are
\begin{align}\label{eq:spin_boson_relation}
S^-_{\bq}
&=
\sqrt S
\left(
a^\dagger_{-\bq}
+
b_{\bq}
\right),
\nonumber\\
S^+_{\bq}
&=
\sqrt S
\left(
a_{\bq}
+
b^\dagger_{-\bq}
\right).
\end{align}

Up to an additive constant, the quadratic spin-wave Hamiltonian is
\begin{equation}
H_2
=
\sum_{\bq}
\begin{pmatrix}
a_{\bq}^\dagger & b_{-\bq}
\end{pmatrix}
\begin{pmatrix}
\gamma_{1,\bq}+\gamma_{2,\bq}
&
\gamma_{3,\bq}
\\
\gamma_{3,\bq}
&
-\gamma_{1,\bq}+\gamma_{2,\bq}
\end{pmatrix}
\begin{pmatrix}
a_{\bq}
\\
b_{-\bq}^\dagger
\end{pmatrix}.
\label{eq:altmag_H2_matrix}
\end{equation}
The coefficients are
\begin{align}
\gamma_{1,\bq}
&=
S(J_{\rm nm}-J_{\rm d})
\left[
\cos(2q_x)-\cos(2q_y)
\right],
\label{eq:altmag_gamma1}
\\
\gamma_{2,\bq}
&=
S(J_{\rm nm}+J_{\rm d})
\left[
\cos(2q_x)+\cos(2q_y)-2
\right]
+
4SJ,
\label{eq:altmag_gamma2}
\\
\gamma_{3,\bq}
&=
4SJ \cos q_x \cos q_y .
\label{eq:altmag_gamma3}
\end{align}

The Bogoliubov transformation
\begin{equation}
\begin{pmatrix}
a_{\bq}
\\
b_{-\bq}^\dagger
\end{pmatrix}
=
\begin{pmatrix}
u_{\bq} & v_{\bq}
\\
v_{\bq}^* & u_{\bq}^*
\end{pmatrix}
\begin{pmatrix}
\alpha_{\bq}
\\
\beta_{-\bq}^\dagger
\end{pmatrix},
\label{eq:altmag_Bogoliubov}
\end{equation}
with $|u_{\bq}|^2-|v_{\bq}|^2=1$, diagonalizes Eq.~\eqref{eq:altmag_H2_matrix}.
Since the coefficients in Eq.~\eqref{eq:altmag_H2_matrix} are real
and even in momentum, the Bogoliubov amplitudes can be chosen real
and even away from exact zero modes.
The resulting magnon
frequencies are~\cite{BrekkeBrataasSudbo2023}
\begin{equation}
\omega_{\alpha,\bq}
=
\Omega_{\bq}
+
\gamma_{1,\bq},
\qquad
\omega_{\beta,\bq}
=
\Omega_{\bq}
-
\gamma_{1,\bq},
\label{eq:altmag_magnon_energies}
\end{equation}
with
\begin{equation}
\Omega_{\bq}
=
\sqrt{
\gamma_{2,\bq}^2
-
\gamma_{3,\bq}^2
}.
\end{equation}
The branch-energy difference
\(\omega_{\alpha,\bq}-\omega_{\beta,\bq}
=2\gamma_{1,\bq}
\propto\cos(2q_x)-\cos(2q_y)\) has \(d_{x^2-y^2}\) symmetry and vanishes on the diagonal
lines \(q_x=\pm q_y\).

Using Eqs.~\eqref{eq:spin_boson_relation} and \eqref{eq:altmag_Bogoliubov} in this gauge, we obtain
\begin{align}
S^-_{\bq}|0\rangle
&=
\sqrt S
\left(
u_{\bq}
+
v_{\bq}
\right)
\alpha^\dagger_{-\bq}|0\rangle,
\nonumber\\
S^+_{\bq}|0\rangle
&=
\sqrt S
\left(
u_{\bq}
+
v_{\bq}
\right)
\beta^\dagger_{-\bq}|0\rangle,
\end{align}
where \(|0\rangle\) is the Bogoliubov vacuum.
Hence, using the evenness of the magnon frequencies, the one-magnon
transverse structure factors are
\begin{align}
S^{+-}_{1{\rm mag}}(\bq,\omega)
&=
S F_{\bq}\,
\delta(\omega-\omega_{\alpha,\bq}),
\label{eq:altmag_Spm}
\\
S^{-+}_{1{\rm mag}}(\bq,\omega)
&=
S F_{\bq}\,
\delta(\omega-\omega_{\beta,\bq}),
\label{eq:altmag_Smp}
\end{align}
where
$F_{\bq}
=
(u_{\bq}+v_{\bq})^2$ is the common one-magnon form factor.
For the stable positive-energy branch,
\[
u_{\bq}^2+v_{\bq}^2
=
\frac{\gamma_{2,\bq}}{\Omega_{\bq}},
\qquad
2u_{\bq}v_{\bq}
=
-\frac{\gamma_{3,\bq}}{\Omega_{\bq}},
\]
and hence
\begin{equation}
F_{\bq}
=
\sqrt{
\frac{
\gamma_{2,\bq}-\gamma_{3,\bq}
}{
\gamma_{2,\bq}+\gamma_{3,\bq}
}
}.
\label{eq:altmag_form_factor_app}
\end{equation}
Expressions at exact zero modes are understood as limits in momentum space.
Using
\(C^{xy}=(S^{+-}-S^{-+})/4\), we obtain
\begin{align}
C^{xy}_{1{\rm mag}}(\bq,\omega)
&=
\frac{S}{4}F_{\bq}
\left[
\delta(\omega-\omega_{\alpha,\bq})
-
\delta(\omega-\omega_{\beta,\bq})
\right].
\label{eq:altmag_Cxy_app}
\end{align}

For small \(\lvert\gamma_{1,\bq}\rvert\), expanding the two peaks in
Eq.~\eqref{eq:altmag_Cxy_app} about \(\Omega_{\bq}\) gives, in the
sense of distributions,
\begin{equation}
C^{xy}_{1{\rm mag}}(\bq,\omega)
=
-\frac{S}{2}F_{\bq}\gamma_{1,\bq}
\delta'(\omega-\Omega_{\bq})
+
O(\gamma_{1,\bq}^3).
\label{eq:Cxy_alter_delta}
\end{equation}
The derivative form reflects the cancellation of the two circularly
polarized branches in the degenerate limit.  It also makes the
\(d_{x^2-y^2}\)-type momentum structure manifest: \(F_{\bq}\) and
\(\Omega_{\bq}\) are invariant under \(q_x\leftrightarrow q_y\),
whereas \(\gamma_{1,\bq}\) changes sign.

\section{Linear spin-wave calculation for the triangular-lattice helical magnet}
\label{sec:LSWT_helical_append}

We derive the one-magnon contribution to the
antisymmetric DSF for the
triangular-lattice helical magnet discussed in Sec.~\ref{sec:LSWT_helical}.
We consider the spatially anisotropic triangular-lattice Heisenberg
antiferromagnet in Eq.~\eqref{eq:H_helical}. For \(0<J'<2J\), we choose
\begin{equation}
\mathbf Q=(Q,0),
\qquad
\cos\frac{Q}{2}
=
-\frac{J'}{2J}.
\end{equation}

We introduce a local rotating frame
\((T^x_{\mathbf r},T^y_{\mathbf r},T^z_{\mathbf r})\) by
\begin{align}
S^x_{\mathbf r}
&=
\cos(\mathbf Q\cdot\mathbf r)\,T^z_{\mathbf r}
-
\sin(\mathbf Q\cdot\mathbf r)\,T^x_{\mathbf r},
\label{eq:helix_local_frame_x}
\\
S^y_{\mathbf r}
&=
\sin(\mathbf Q\cdot\mathbf r)\,T^z_{\mathbf r}
+
\cos(\mathbf Q\cdot\mathbf r)\,T^x_{\mathbf r},
\label{eq:helix_local_frame_y}
\\
S^z_{\mathbf r}
&=
T^y_{\mathbf r}.
\label{eq:helix_local_frame_z}
\end{align}
The classical helical state is then mapped to a uniformly polarized state in the rotating frame,
\begin{equation}
\langle T^z_{\mathbf r}\rangle=S .
\end{equation}
This is the standard rotating-frame formulation of spin waves in helical
magnets~\cite{CooperElliottNettelSuhl1962,Kataoka1987}.

We next express the physical spin operators in terms of the rotating-frame
operators.  Let \(N\) denote the number of sites of the triangular lattice.
With the Fourier convention
\(O_{\bq}=N^{-1/2}\sum_{\mathbf r}
e^{-i\bq\cdot\mathbf r}O_{\mathbf r}\),
the one-magnon parts of the physical transverse spin operators are
\begin{align}
\left.S^x_{\bq}\right|_{1{\rm mag}}
&=
-\frac{1}{2i}
\left(
T^x_{\bq-\mathbf Q}
-
T^x_{\bq+\mathbf Q}
\right),
\label{eq:helix_Sx_sideband}
\\
\left.S^y_{\bq}\right|_{1{\rm mag}}
&=
\frac12
\left(
T^x_{\bq-\mathbf Q}
+
T^x_{\bq+\mathbf Q}
\right).
\label{eq:helix_Sy_sideband}
\end{align}
Thus the one-magnon response at momentum \(\bq\) contains two
rotating-frame momentum channels, \(\bq\pm\mathbf Q\).

We use the Holstein--Primakoff representation in the rotating frame,
\begin{equation}
T^+_{\mathbf r}
=
\sqrt{2S}\,a_{\mathbf r},
\quad
T^-_{\mathbf r}
=
\sqrt{2S}\,a^\dagger_{\mathbf r},
\quad
T^z_{\mathbf r}
=
S-a^\dagger_{\mathbf r}a_{\mathbf r}.
\end{equation}
The quadratic spin-wave Hamiltonian is
\begin{equation}
H_2
=
\sum_{\bq}
\left[
A_{\bq}a^\dagger_{\bq}a_{\bq}
+
\frac{B_{\bq}}{2}
\left(
a_{\bq}a_{-\bq}
+
a^\dagger_{\bq}a^\dagger_{-\bq}
\right)
\right],
\label{eq:helix_H2}
\end{equation}
where
\begin{align}
A_{\bq}
&=
S\left[
\frac12
\left(
J_{\bq+\mathbf Q}
+
J_{\bq-\mathbf Q}
\right)
+
J_{\bq}
-
2J_{\mathbf Q}
\right],
\label{eq:helix_Ak}
\\
B_{\bq}
&=
S\left[
\frac12
\left(
J_{\bq+\mathbf Q}
+
J_{\bq-\mathbf Q}
\right)
-
J_{\bq}
\right].
\label{eq:helix_Bk}
\end{align}
Here \(J_{\bq}\) is the Fourier transform of the exchange coupling
defined in Eq.~\eqref{eq:Jq_triangular}.

Away from exact zero modes, we use the real, even Bogoliubov transformation
\begin{equation}
a_{\bq}
=
u_{\bq}\alpha_{\bq}
+
v_{\bq}\alpha^\dagger_{-\bq},
\qquad
u_{\bq}^2-v_{\bq}^2=1.
\end{equation}
It diagonalizes Eq.~\eqref{eq:helix_H2} and
gives
\begin{align}
\omega_{\bq}
&=
\sqrt{
A_{\bq}^2
-
B_{\bq}^2
}\nonumber\\
&=
S
\sqrt{
2
\left(
J_{\bq}
-
J_{\mathbf Q}
\right)
\left(
J_{\bq+\mathbf Q}
+
J_{\bq-\mathbf Q}
-
2J_{\mathbf Q}
\right)
}.
\label{eq:helix_omega_J}
\end{align}
For the stable positive-energy branch, the Bogoliubov amplitudes satisfy
\[
u_{\bq}^2+v_{\bq}^2
=
\frac{A_{\bq}}{\omega_{\bq}},
\qquad
2u_{\bq}v_{\bq}
=
-\frac{B_{\bq}}{\omega_{\bq}}.
\]
The behavior at exact zero modes is obtained by taking the
corresponding limits in momentum space.
In the isotropic triangular-lattice limit, Eq.~\eqref{eq:helix_omega_J} reduces to the standard
spin-wave dispersion of the \(120^\circ\)
antiferromagnet~\cite{JolicoeurLeGuillou1989,MomoiSuzuki1992,ChernyshevZhitomirsky2009}.

To leading order,
\[
T^x_{\bk}
=
\sqrt{\frac S2}
\left(
a_{\bk}+a^\dagger_{-\bk}
\right).
\]
We define its one-magnon spectral weight by
\begin{equation}
W_{\bk}\delta(\omega-\omega_{\bk})
\equiv
\sum_{n\in1{\rm mag}}
\langle0|T^x_{-\bk}|n\rangle
\langle n|T^x_{\bk}|0\rangle
\delta(\omega-E_n+E_0),
\label{eq:helix_Tx_spectral_weight}
\end{equation}
where \(|0\rangle\) is the Bogoliubov vacuum.
Using the Bogoliubov transformation, one finds
\begin{equation}
W_{\bk}
=
\frac S2
\left(u_{\bk}+v_{\bk}\right)^2
=
\frac S2
\sqrt{
\frac{
A_{\bk}-B_{\bk}
}{
A_{\bk}+B_{\bk}
}
}
=
S^2
\frac{
J_{\bk}-J_{\mathbf Q}
}{
\omega_{\bk}
}.
\label{eq:helix_Wk_app}
\end{equation}

Substituting Eqs.~\eqref{eq:helix_Sx_sideband} and
\eqref{eq:helix_Sy_sideband} into the Lehmann representation of
\(S^{xy}(\bq,\omega)\), momentum conservation eliminates the cross
terms between the two sidebands.  The remaining one-magnon
contribution is
\begin{equation}
S^{xy}_{1{\rm mag}}(\bq,\omega)
=
\frac{i}{4}
\left[
W_{\bq+\mathbf Q}
\delta(\omega-\omega_{\bq+\mathbf Q})
-
W_{\bq-\mathbf Q}
\delta(\omega-\omega_{\bq-\mathbf Q})
\right].
\label{eq:helix_Sxy_app}
\end{equation}
Equation~\eqref{eq:helix_Sxy_app} is purely imaginary, and hence
\(C^{xy}_{1{\rm mag}}=\operatorname{Im}S^{xy}_{1{\rm mag}}\) gives
\begin{equation}
C^{xy}_{1{\rm mag}}(\bq,\omega)
=
\frac14
\left[
W_{\bq+\mathbf Q}
\delta(\omega-\omega_{\bq+\mathbf Q})
-
W_{\bq-\mathbf Q}
\delta(\omega-\omega_{\bq-\mathbf Q})
\right].
\label{eq:helix_Cxy_LSWT_app}
\end{equation}
This expression reproduces Eq.~\eqref{eq:helix_Cxy_LSWT} in the main text.

\bigskip
\bigskip

\bibliography{dynamical_structure_factor}

\end{document}